\documentclass{article}
\usepackage{emulateapj}
\usepackage{onecolfloat}
\usepackage{graphicx}

\lefthead{Weiner {\it et al.}}
\righthead{Disk and Halo Mass of NGC 4123. I.}

\slugcomment{\it Accepted to ApJ}

\long\def\Ignore#1{\relax}

\newcommand\captsize{\scriptsize}

\newcommand\kms{\hbox{$\rm{km}~\rm{s}^{-1}$}}
\newcommand\ha{\hbox{H$\alpha$}}
\newcommand\hi{\hbox{H~I}} 
\newcommand\mden{\hbox{${\rm M}_{\odot}~{\rm pc^{-3}}$}}

\newcommand\percmsq{\hbox{${\rm cm}^{-2}$}}

\def\deg{\hbox{$^{\circ}$}} 
\newcommand\msun{\hbox{${\rm M}_{\odot}$}}
\newcommand\lsun{\hbox{${\rm L}_{\odot}$}}
 
\newcommand\etal{{\it et al.}}
\newcommand\eg{{\it e.g.}}
\newcommand\ml{\hbox{$M/L$}}

\def\hii{H\,\,\textsc{ii}}
\def\lsun{\hbox{$L_{\odot}$}}

\def\mli{\hbox{$M/L_I$}}

\def\chisq{\hbox{$\chi^2$}}
\def\rchisq{\hbox{$\chi^2/N$}}

\def\beginfig{\begin{figure}}
\def\endfig{\end{figure}}
\def\beginfigtwo{\begin{figure*}}
\def\endfigtwo{\end{figure*}}

\def\begintab{\begin{table}}
\def\endtab{\end{table}}
\def\begintabtwo{\begin{table*}}
\def\endtabtwo{\end{table*}}

\begin{document}

\twocolumn[

\title{The Disk and Dark Halo Mass of the Barred Galaxy NGC 4123.\\
I. Observations}

\author{Benjamin J.\ Weiner\altaffilmark{1,2}}
\affil{Observatories of the Carnegie Institution of Washington,
813 Santa Barbara St, Pasadena, CA  91101}

\author{T.\ B.\ Williams\altaffilmark{1}}
\affil{Department of Physics and Astronomy, Rutgers
University, 136 Frelinghuysen Rd., Piscataway, NJ 08854}

\author{J.\ H.\ van Gorkom\altaffilmark{2}}
\affil{Department of Astronomy, Columbia University,
538 W. 120th St., New York, NY  10227}

\author{and}
\author{J.\ A.\ Sellwood}
\affil{Department of Physics and Astronomy, Rutgers
University, 136 Frelinghuysen Rd., Piscataway, NJ 08854}


\begin{abstract}
The non-circular streaming motions in barred galaxies are sensitive to 
the mass of the bar and can be used to lift the degeneracy between
disk and dark matter halo encountered when fitting axisymmetric 
rotation curves of disk galaxies.  
In this paper, we present photometric and kinematic observations of
NGC 4123, a barred galaxy of modest size ($V_{rot} = 130$ \kms,
$L = 0.7 L_*$), which reveal strong non-circular motions.  The
bar has straight dust lanes and an inner Lindblad resonance.  The disk
of NGC 4123 has no sign of truncation out to 10 scale lengths, and
star-forming regions are found well outside $R_{25}$.  A Fabry-Perot
\ha\ velocity field shows velocity jumps of $>100$ \kms\ at the
location of the dust lanes within the bar, indicating shocks in the
gas flow.  VLA observations yield the velocity field of the 
\hi\ disk.  Axisymmetric mass models yield good fits to the rotation 
curve outside the bar regionfor disk $I$-band \ml\ of 2.25 or less,
and dark halos with either isothermal or power-law profiles can fit
the data well.  In a companion paper, we model the full 2-D velocity
field, including non-circular motions, to determine the stellar \ml\
and the mass of the dark halo.

\end{abstract}


 ]

\altaffiltext{1}{Visiting Astronomer, Cerro Tololo Inter-American 
Observatory.}

\altaffiltext{2}{Visiting Astronomer, National Radio Astronomy 
Observatory.}


\section{Introduction}

The rotation speed of gas in a disk galaxy is directly related
to its mass, providing a dynamical measure of galaxy mass 
if the gas is on circular orbits in the disk plane
(\eg\ \"Opik 1922; Roberts 1969; Tully \& Fisher 1977).
The rotation curve $V_c(R)$
yields the radial profile of the enclosed mass $M(<R)$, modulo
the effects of the flattening of the mass distribution.
Many galaxies have nearly flat rotation curves, requiring
dark mass more extended than the luminous material
(\eg\ Babcock 1939; Rubin, Ford, \& Thonnard 1978; Bosma 1981).

Unfortunately, even a high quality rotation curve extending well 
beyond the optical disk does not constrain the relative contributions of 
dark and visible matter to the centripetal acceleration; 
such decompositions are almost totally degenerate.  For example, 
van Albada \etal\ (1985) showed that the rotation curve of the 
well-observed galaxy NGC 3198, the ``poster child'' for dark matter 
(DM) halos, can be fitted equally well with a massive disk and a dark halo 
significant only in the outer parts of the galaxy, a less massive disk and 
a more dominant halo, or even just a halo and a zero-mass disk.

It is critical to try to determine the relative masses of disks and halos 
in galaxies because uncertainty in this quantity impacts almost every area 
of galaxy formation and dynamics.  It hampers our understanding of: the 
structure and formation process of disk galaxies; the efficiency
of assembly of baryons into disks;
the Tully-Fisher relation between luminosity and rotation width;
instabilities which give rise to 
spirals and bars, warps, mergers, etc. Furthermore, 
ignorance of dark halo densities and masses leaves
us without a probe of the initial density 
fluctuation power spectrum on galaxy scales.  

One approach to this problem has been to place a lower bound 
on the mass of the inner DM halo by constructing a ``maximum disk'' model 
to fit the rotation curve (Kalnajs 1983; van Albada \& Sancisi 1986).  
Such a model attributes the highest possible 
mass-to-light ratio (\ml) to the luminous matter, usually consistent with 
a non-hollow DM halo.  There are a number of indirect dynamical arguments 
suggesting that the disk mass is in fact close to maximum for most bright 
galaxies (\eg\ van Albada \& Sancisi 1986; Freeman 1992; Debattista \& 
Sellwood 1998) or nearly so 
(Athanassoula, Bosma \& Papaioannou 1987; Bosma 1999).  
Counter-arguments against maximum disks have been advanced by 
Efstathiou, Lake, \& Negroponte (1982), Bahcall \& 
Casertano (1985), Bottema (1993, 1997), van der Kruit (1995), Courteau \& 
Rix (1999) and others.  Several recent studies, motivated by cosmological 
structure formation simulations, have addressed dark halo structure and 
the relation between dark halo size and galaxy size (\eg\ Navarro \etal\ 
1996, Syer \& White 1998, Syer, Mao, \& Mo 1999) and have generally 
favored strongly sub-maximal disk models. 

While models with zero mass disks can fit rotation curves,
all reasonable models attribute some mass to the luminous component.
The disagreement between maximum disk and halo-dominated models such
as those of Bottema (1997) is roughly a factor of 2.5 in disk \ml.
Population synthesis estimates of stellar \ml\ ratios
cannot resolve this discrepancy, because most of the light is
emitted by massive stars while most of the mass is in faint stars
(Worthey 1994; Charlot, Worthey \& Bressan 1996).

Barred galaxies offer a way to break the disk-halo degeneracy.  The 
elongated bar drives non-circular streaming motions in stars and gas 
within the bar region.  The magnitude of the non-circular motions depends 
mostly on the mass of the bar and, to a lesser extent, on the pattern 
speed at which it rotates.  The departures from axisymmetry in both light 
distribution and kinematics can be observed.  The (unobservable) dark halo 
is thought to be mostly ``pressure'' supported and therefore more rounded 
than the bar.  The stellar
bar is therefore the dominant source of non-circular
streaming motions, which allow us to estimate its \ml, and by extension 
that of the stellar disk.

In practice, we need extensive observations and modeling to carry out this
program.  The most easily observable kinematic tracers are optical and radio 
emission lines from gas in the galaxy disk.  Because the gas is not on 
circular orbits, and we observe only line-of-sight velocities, we cannot 
infer the full space velocity field of the gas, as we can for an 
axisymmetric galaxy.  
Furthermore, gas flow in a barred potential 
generally gives rise to shocks, which prevent us from directly inverting
the observed gas velocities to derive the mass distribution.  
We therefore need to derive the bar mass by matching fluid models
incorporating shocks to the observed velocity field.  


In this paper we describe a detailed study of the barred galaxy NGC 4123.
We selected this galaxy for its fairly strong bar, a moderate 
inclination, favorable projection angle of the bar, 
colors and morphology suggesting substantial \ha\ emission,
and modest size. 
Because projection of a non-axisymmetric velocity field into the 
observable line-of-sight velocity depends on its angle to the galaxy major 
axis (line of nodes), candidates for study must be selected carefully 
(Pence \& Blackman 1984b; Long 1991).  
If the bar is aligned with either the major or minor axis of the galaxy disk
as it appears on the sky  (\eg\ NGC 1300, England 1989), it is 
difficult to separate the circular and non-circular parts of the 
velocity field after projection, since the orbits or streamlines are
elongated in approximately the same direction as the bar
(\eg\ Athanassoula 1992b).  
The projection is most favorable for bars at intermediate position angles 
-- a 45\deg\ angle between the bar and the line of nodes is best. 

We require two-dimensional surface photometry in several broad-band 
filters and a 2-D velocity map over the entire galaxy.  
``Three-dimensional'' imaging spectroscopy is possible with Fabry-Perot 
interferometers at optical wavelengths, and with aperture-synthesis 
interferometers at radio wavelengths.  These techniques are complementary
in angular resolution and field of view, and the spectral lines 
arise from different phases of the interstellar medium.
This galaxy is the first of several barred galaxies we have observed with
the Rutgers Imaging Fabry-Perot Interferometer at CTIO 
(Schommer \etal\ 1993).

We present surface photometry and velocity maps of NGC 4123,
from which we construct axisymmetric 
models.  Mass models with a maximum disk \ml\ of 2.25 and either isothermal
or power-law profiles can fit the rotation curve well.
In a companion paper (Weiner \etal\ 2000, hereafter Paper 
II), we model the full 2-D velocity field to determine the separate masses 
of the disk and halo.

\section{NGC 4123 -- Basic properties}

NGC 4123 is of moderate size and has no pronounced morphological 
peculiarities.  It is an SB(r)c galaxy with total blue magnitude 
$B_T = 11.98$, a rotation amplitude of approximately 130~\kms, and a 
recession velocity of $1327~\kms$, from the RC3 catalog 
(de Vaucouleurs \etal\ 1991), via 
NED\footnote{The NASA/IPAC Extragalactic Database (NED) is operated by the 
Jet Propulsion Laboratory, California Institute of Technology, under 
contract with the National Aeronautics and Space Administration.}.  
We adopt a distance of 22.4 Mpc ($m-M=31.75$), based on a Hubble constant of 
75 \kms~Mpc$^{-1}$ and a correction to the microwave background rest frame,
with no Virgocentric correction.  Dynamical estimates of \ml\ vary inversely 
with the adopted distance $D$; \ml\ 
values given in this paper are in solar units and 
are implicitly followed by the factor 
$h_{75}$ ($H_0 / 75~\kms~\rm{Mpc}^{-1}$).  
However, estimates of the degree of disk maximality (disk-to-halo ratio)
do not depend on adopted distance.  The basic properties of NGC 4123
are summarized in Table \ref{table-basicprops}.


\begintab[t]
\begin{center}
\begin{tabular}{ll}
\tableline\tableline 

Parameter & Value  \\ 
\tableline

Recession velocity & 1327 \kms \\
Assumed distance &  22.4 Mpc \\
$m_B$ (within $3R_{25}$)  & 12.06 \\
$m_V$ (within $3R_{25}$)  & 11.45 \\
$m_I$ (within $3R_{25}$)  & 10.33 \\
PA      & $-33 \pm 2$\deg \\
inclination	& $45 \pm 4$ \deg \\
$L_I$ (extinction corrected)  & $1.72 \times 10^{10}~\lsun$ \\
$V_{rot}$ & 130 \kms \\
21 cm flux & 63.5 Jy \kms \\
$M_{H I}$ & $7.5 \times 10^9~\msun$ \\

\tableline
\end{tabular}
\end{center}
\caption{NGC 4123 -- Basic Properties}
\label{table-basicprops}
\endtab

NGC 4123 has a fairly strong bar, but the bar does not totally dominate 
the disk (unlike \eg\ NGC 1097 or NGC 1365; cf. Ondrechen \& van der Hulst 
1989; Ondrechen \etal\ 1989; Lindblad, Lindblad, \& Athanassoula 1996),
so it is not an extreme case.
The galaxy has a favorable inclination (45\deg\ as 
determined from 21 cm observations, see below) and the bar is aligned at 
approximately 53\deg\ to the line of nodes when deprojected, which is 
almost ideal for recovering non-circular motions.

NGC 4123 has a companion galaxy, NGC 4116, which lies 
14\arcmin\ southwest of NGC 4123 and at nearly the same systemic 
velocity, $1312~ \kms$.  NGC 4116 is of SB(rs)dm type and is fainter than 
NGC 4123, with magnitude $B_T = 12.41$, from the RC3 via NED.  At 
22.4 Mpc, the separation of the galaxies in the plane of the sky
is just 91 kpc.  However, neither shows much sign of tidal interaction 
(see Section \ref{sec-vlaobs}) and it is possible that the real separation 
is much greater.  The largely undisturbed kinematics of the 
two galaxies, shown in Taylor \etal\ (1995) and discussed further below, 
indicate that any interaction does not visibly affect their 
internal dynamics.  The conclusions we draw from studying the dynamics of 
the bar region of NGC 4123 are unaffected by the presence of NGC 4116.

\section{Surface Photometry}

We require surface photometry in order to estimate the distribution of 
luminous mass in the galaxy.  It is usual to assume that mass is 
proportional to luminosity; a uniform \ml\ has the advantage of being the 
simplest possible assumption, and is generally thought to be reasonable 
when color variations are small.  This is difficult to prove,
since the bulk of the mass is in low mass stars while high mass stars emit 
much more light per unit mass.  Furthermore, extinction is particularly 
variable in barred galaxies, which generally have strong dust lanes within 
the bar.  Such problems are lessened at red and near IR wavelengths, since 
bluer optical bands are more strongly affected by emission from young 
massive stars and by spatially variable extinction, although $I$ (and even 
$K$) band observations are still affected by young stars and by dust
(de Jong 1996b, Rhoads 1998).

\subsection{Observations}

The initial imaging observations of NGC 4123 were made on the night of 
1994 April 1 with the 0.9-meter telescope of Cerro Tololo Interamerican 
Observatory.\footnote{CTIO is operated by the Association of
Universities for Research in Astronomy, Inc.\ under a cooperative
agreement with the National Science Foundation.}
We used a Tektronix 1K x 1K CCD at the f/13.5 Cassegrain focus, with a 
pixel scale of 0.39\arcsec, and a field of view of 6.6\arcmin.  
We observed NGC 4123 in $BVRI$ filters,
obtaining 10 minute exposures in $BVR$ and $3 \times 10$ minute exposures
in $I$, offset between exposures to improve flatness of the final
image.  
The seeing was 1.2\arcsec\ in $VRI$, or 130 pc at the adopted
distance of 22.4 Mpc, and 1.4\arcsec\ in $B$.  
We reduced the images with IRAF\footnote{IRAF is distributed 
by NOAO, which is operated by AURA
Inc., under a cooperative agreement with the National Science
Foundation.}, subtracting
overscan and a bias frame made by a robust combination
of many individual biases, and flatfielding with twilight flats
in the usual way.  Photometric standard stars from the E-regions
(Graham 1982) were observed at several times during the night
and used to derive extinction coefficients.

The field of view of the CTIO observations is rather small for
such a large, nearby galaxy.  We reobserved NGC 4123 to obtain a
larger field, using the Swope 1-meter telescope at 
Las Campanas Observatory on the night of 1998 January 25.
These observations used the LCO Tek\#5 2K x 2K CCD at the
f/7 Cassegrain focus, with a
pixel scale of 0.70\arcsec\ and field of view of 24\arcmin.
We obtained 10 minute exposures of NGC 4123 in $BV$ filters
and $2 \times 10$ minute offset exposures in the $I$ filter.  
The seeing was approximately 1.6\arcsec\ in $VI$ and 1.8\arcsec\
in $B$.  The images were again reduced with IRAF, subtracting
overscan and bias and flatfielding with twilight flats.  
Photometric standard stars from Landolt (1992) were observed throughout
the night and used to derive extinction coefficients.
The photometric calibrations of the CTIO and LCO data are consistent.

The LCO Tek\#5 chip exhibits fringing and illumination gradients in the 
$I$ filter which cannot be removed with twilight flats, since the
night sky spectrum is of a different color and contains strong emission
lines.  We constructed
a supersky image by combining a number of dithered $I$ images
from throughout the 3-night run, scaled by the sky brightness,
and subtracted this image from the NGC 4123 $I$ image.  This process
removed both small-scale fringes and large-scale gradients in the image.

\subsection{Morphology}

\beginfigtwo[t]
\begin{center}
\includegraphics[width=5.75truein]{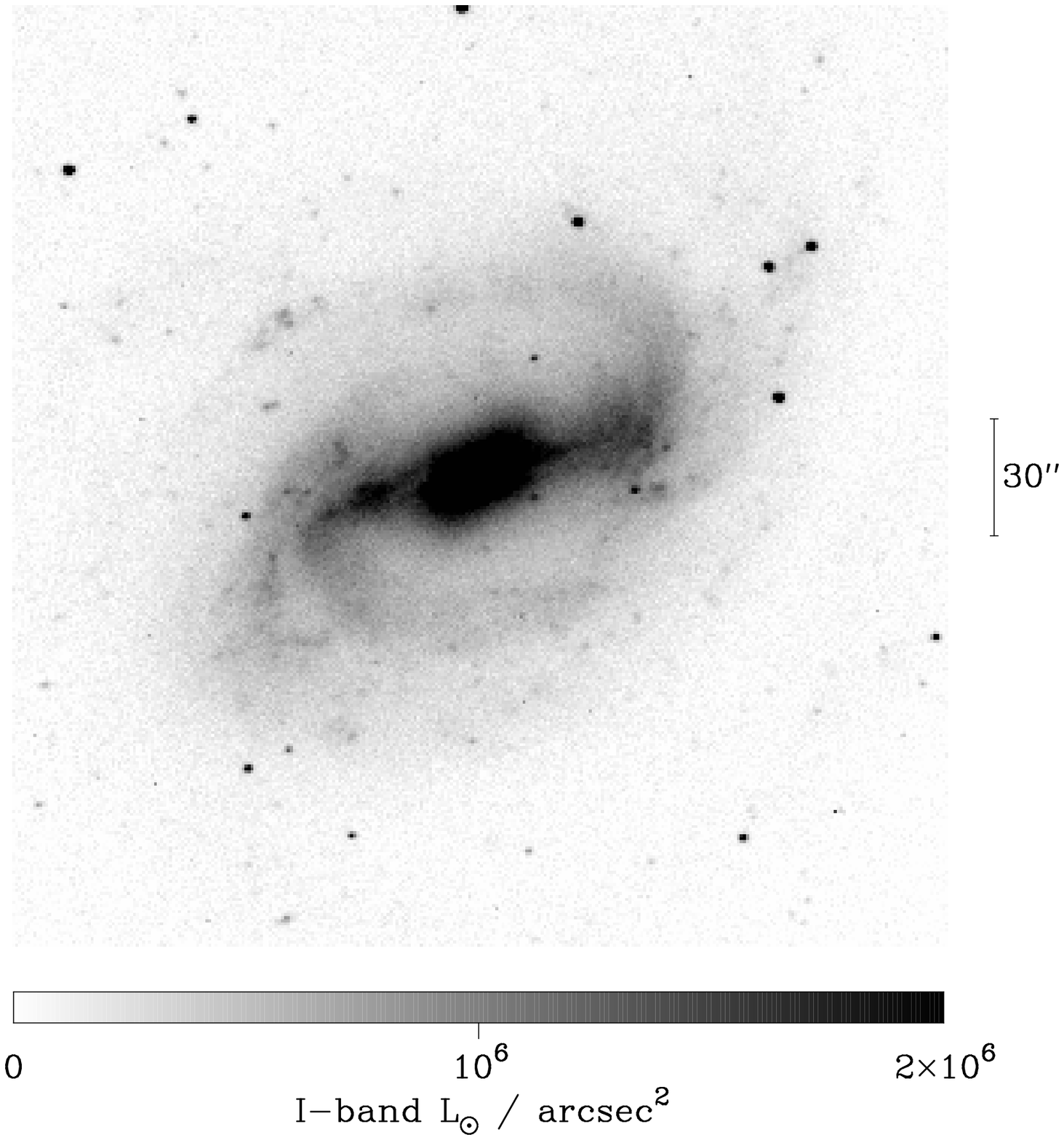}
\end{center}

\caption{NGC 4123, $I$ image.}
\captsize
The inner 240\arcsec\ square region of the CTIO 0.9-m $I$-band image of 
NGC 4123, on a linear scale.  North is up and west is to the right.  The 
major axis of the disk is 33\deg\ west of north.  
[Resolution degraded by a factor of 2 for preprint version.]
\label{fig-iimage}
\endfigtwo

The inner part of the combined CTIO 0.9-m $I$ image of NGC 4123 is 
presented in Figure \ref{fig-iimage}.  
The bar has a semi-major axis of approximately 50\arcsec\ on the sky,
as determined by the transition from boxy to essentially
elliptical isophotes, and by the break in the surface brightness
profile (see below),
or 5.6 kpc when deprojected.
The bar has a rather complex shape, with a bright boxy 
component near the center and a fainter elongated component, which is 
of constant width or even broadens a little towards the ends;
there is some evidence for an isophotal twist between the two
components.
The complex shape makes it difficult to define a single axis ratio for the 
bar: the elongated component has an axis ratio of about 6:1, while the boxy 
component has an axis ratio of about 2.5:1.

Other than the central source (discussed below), NGC 4123 has no 
significant spherical bulge, which makes it particularly suitable for the 
deprojection method we use in our simulations of the gas flow in this 
galaxy (see Paper II).
Outside the bar region, the disk of NGC 4123 is relatively 
smooth in $I$ light, with some spiral structure but no sharp features or
strong brightness variations.  

There is an isophotal twist in the outer regions of the disk.  At very 
large radii the PA and axis ratio of the isophotes are consistent with the 
PA of the galaxy major axis and the inclination as derived from 21 cm 
observations of the \hi\ disk of the galaxy (see section 
\ref{sec-vlaobs}).  However, at radii less than $\sim 100\arcsec$, but 
well outside the bar, the PA of the disk isophotes is intermediate between 
the PA of the bar and the PA of the kinematic major axis indicated by the 
21 cm observations.  
A heavily smoothed image shows a faint spiral arm pattern
in the outer disk, which is responsible for the isophotal twist.
There is no large-scale twist in the kinematic line of nodes in this 
range of radii in the \hi\ observations, which indicates that the 
isophotal twist is not accompanied by a warp (Briggs 1990).

\beginfigtwo[t]
\begin{center}
\includegraphics[width=5.75truein]{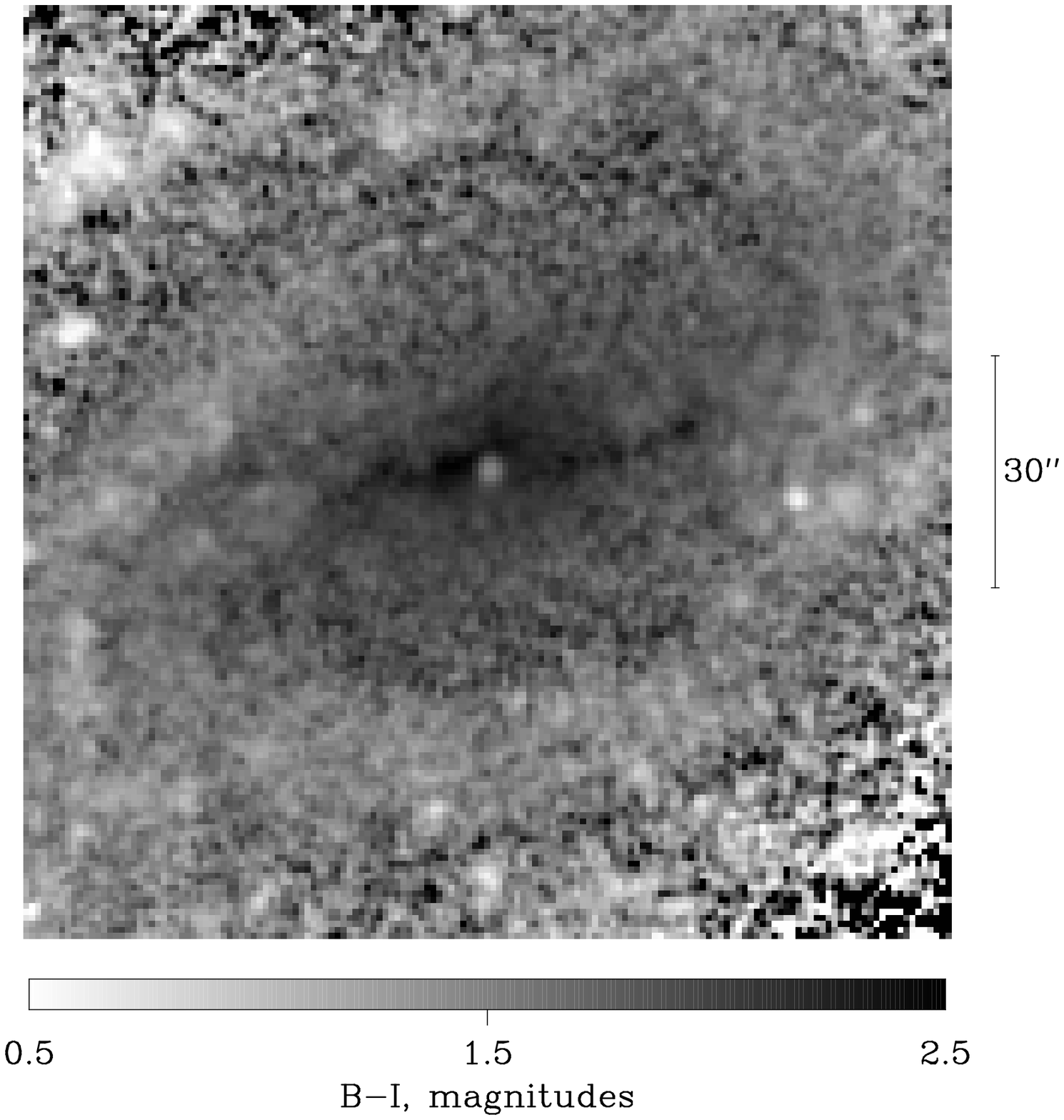}
\end{center}

\caption{NGC 4123, $B-I$ color map}
\captsize
The inner 120\arcsec\ square region of the CTIO $B-I$ color map of NGC 
4123.  North is up and west is to the right.  Darker colors are redder.  
The dust lanes in the bar are apparent as the dark lines running east and 
west of the nucleus.  The nucleus of the galaxy is much bluer than its 
surroundings.  The blue spots at the west end of the bar are foreground 
stars.
[Resolution degraded by a factor of 2 for preprint version.]
\label{fig-colormap}
\endfigtwo

The $B-I$ color map, presented in Figure \ref{fig-colormap}, was 
constructed from CTIO $B$ and $I$ images convolved to 1.8\arcsec\ 
resolution.  The dust lanes, conspicuous in this color map,
are linear features, offset from the bar major axis towards its 
leading edge.  
Dust lanes of this geometry are common in strongly barred galaxies (Sandage 
1961; Sandage \& Bedke 1988; Athanassoula 1992b and references therein).
Color-color plots of pixels within the bar (as in
Quillen \etal\ 1995) show that in the 
east dust lane, the slopes of $R-I$ and $V-I$ versus $B-I$ are 
consistent with a dust absorbing screen with $R_V = 3.1$
(Cardelli, Clayton \& Mathis 1989), confirming that the dust
lane in the east half is near to us.  In the west dust lane
the slopes are shallower, so that the dust lane is on the
far side of the bar, as expected.

The small bright nucleus at the center of the galaxy stands out as the 
central blue spot.  It is barely 
resolved by these observations.  Its color, $B-I = 1.4$, is quite blue 
compared to its surroundings ($B-I = 2.2$), and its luminosity in the $I$ 
band is $1.8 \times 10^8 \lsun$.  The center of NGC 4123 appears in lists 
of known emission line and UV-excess sources as UM 477 (MacAlpine
\& Williams 1981) and Markarian 1466, suggesting that it is a 
nuclear starburst.
It is likely, therefore, that the \ml\ of the nucleus 
is lower than the overall \ml\ of the stellar disk, a point we return to 
in the models of Paper II.

\subsection{Surface brightness profiles}
\label{sec-sbprofiles}

We constructed axisymmetrized surface brightness profiles of NGC 4123 from 
the Las Campanas $BVI$ data.  The 21\arcmin\ field allows us to extend the 
surface brightness profiles to large galactic radius.

We first masked approximately 1100 stars and faint galaxies and a 
small number of cosmic rays out of the image, 
using large masks around the stars to exclude the faint halo of the stellar 
PSF.  The surface brightness profiles are then constructed
by dividing the image into elliptical annuli, 1.7\arcsec\ wide along the 
major axis, with the disk inclination of 45\deg\ and position angle of 
--33\deg\ as derived from the 21cm \hi\ observations (see below).  In each 
annulus, we compute the trimmed mean with a 2-sigma clipping.  
We determined the sky level by computing the biweight, a robust estimate 
of the mode (Beers, Flynn \& Gebhardt 1990), of all pixels within an 
elliptical annulus of the same inclination and PA, an inner major axis 
radius of 551\arcsec\ (60 kpc) and outer major axis radius of 
593\arcsec\ (65 kpc).
The error estimate is computed directly from the dispersion of the pixel 
values in each annulus, excluding clipped pixels, with the statistical 
error of the sky value added in quadrature.

The axisymmetrized $B, V$, and $I$ surface brightness profiles of NGC 4123,
and the color profiles,
are shown in Figure \ref{fig-sbprofile}.
We have not included a correction for internal extinction in 
these plots, but we have corrected for geometric inclination to
obtain a face-on surface brightness:
\begin{equation}
\mu_{\rm corr} = \mu_{\rm obs} + 2.5~\rm{log} (b/a),
\end{equation}
as appropriate for a transparent galaxy.  At an inclination of 
45\deg\ this correction is $+0.38$ mag~arcsec$^{-2}$.   
The shape of the profiles, the 
colors, and the color gradient are typical of SBbc galaxies.  

There is a substantial color gradient, $\sim 0.4$ mag, inside the bar
radius, but this does not imply an equally large color gradient in the
old stellar population.  Inside the bar, the strong dust lanes produce
reddening, while outside the bar, many \hii\ regions with very blue
colors are present in the spiral arms.  We measured mean colors in
several regions in the bar and disk in and out of dust lanes and \hii\
regions.  The results are given in Table \ref{table-colors}.  When
obvious dust lanes and \hii\ regions are excluded, there
is only an 0.2 mag difference in $B-I$ between the bar and disk,
so there is no evidence for a significant difference in stellar
population.  Since unresolved dust lanes and knots of young 
stars may be present, the actual color difference could be smaller.


\begintab[t]
\begin{center}
\begin{tabular}{ll}
\tableline\tableline 

Region & $B-I$ (mag) \\
\tableline

Dust lane in bar         & 2.3 \\
Bar/lens outside dust lane  & 2.0 \\
Disk, interarm region  & 1.8 \\
Disk, spiral arms  & 1.5  \\

\tableline
\end{tabular}
\end{center}
\caption{NGC 4123 -- Mean Colors}
\label{table-colors}
\endtab

In our subsequent use of the $I$ data to fit the rotation curve and
model the potential, we apply a $-0.15$ mag overall
correction for internal 
extinction in the $I$ band,
calculated from the $A_{\rm int} = -1.0 \log(b/a)$ prescription given by 
Giovanelli \etal\ (1994), which Palunas (1996) found to be reliable for 
similar $I$-band data.  Galactic extinction $A_{\rm ext}$ towards NGC 4123 
is negligible (Burstein \& Heiles 1982).

We did not attempt to correct the $I$ image for differential extinction 
within the galaxy.  This procedure requires extensive multicolor data, 
including near-IR imaging, and detailed modeling of radiative transfer 
(see Byun \etal\ 1994; Regan \etal\ 1995; de Jong 1996b).  To estimate the 
extinction in the east (stronger) dust lane, we selected a region of the bar 
where the dust lane appeared strong in $B$ and took a cut through the 
data perpendicular to the dust lane.  Assuming the unextincted bar light 
profile is symmetric when reflected about the bar major axis, we estimate 
the differential extinction to be $\sim 25\%$ (0.3 mag) in $B$ and 
$\sim 10\%$ (0.1 mag) in $I$.  
Alternatively, for the dust lane reddening of $E_{B-I} = 0.3$,
an $R_V = 3.1$ extinction law (Cardelli, Clayton \& Mathis 1989),
and an absorbing dust screen, the dust lane extinction in $I$ is 0.17 mag.
Since the dust lanes cover only a small part of the bar, the amount 
of $I$ light lost to extinction is just a few percent.

\beginfig[ht]

\begin{center}
\includegraphics[width=3.5truein]{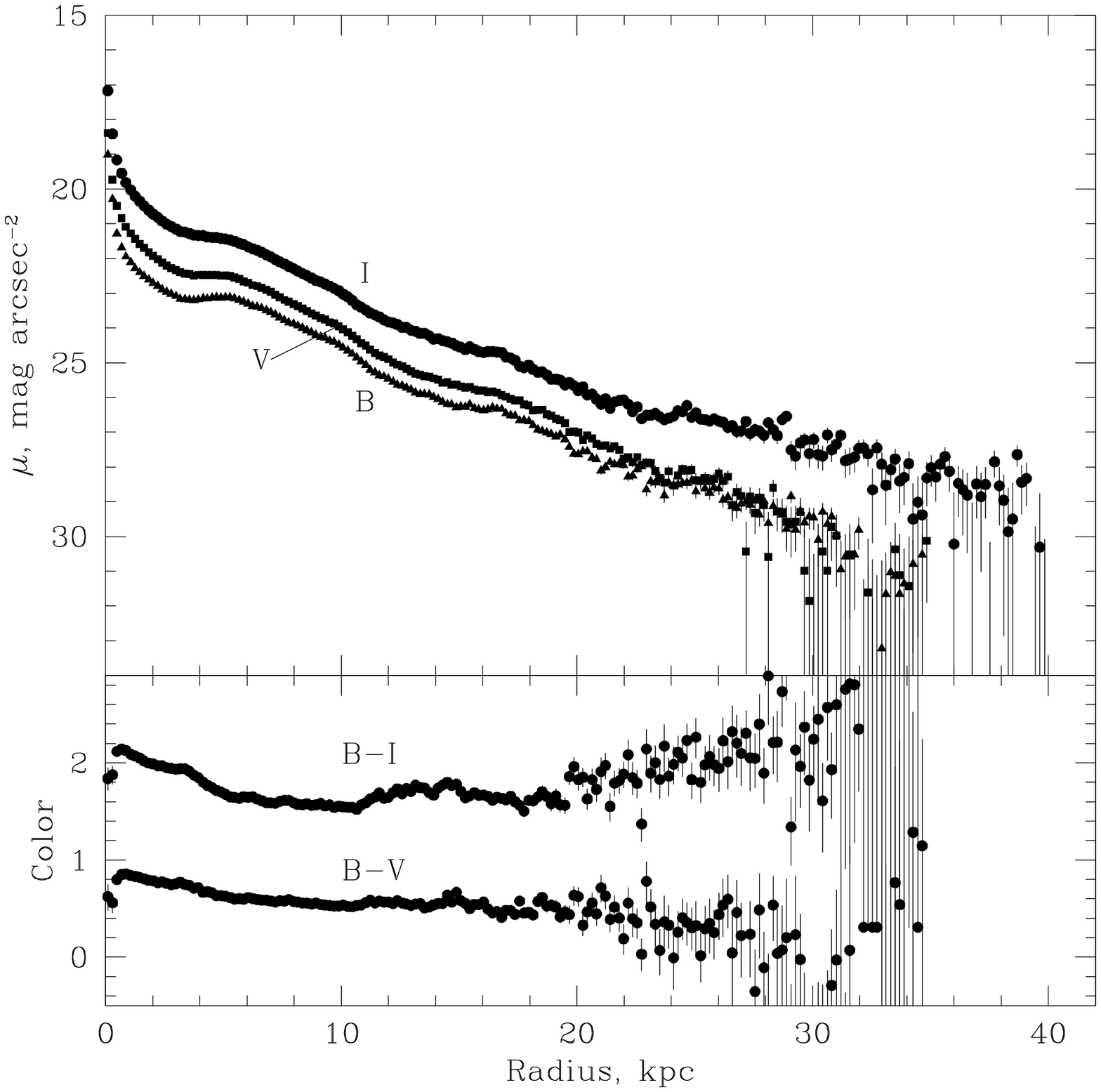}
\end{center}

\caption {$BVI$ surface brightness and color profiles of NGC 4123}
\captsize
Upper panel:
The axisymmetrized LCO $B,V,I$-band surface brightness profiles of NGC 4123, 
in magnitudes arcsec$^{-2}$.  The profiles have been corrected for 
inclination but not internal extinction.  $B,V,I$ are triangles, squares, 
and circles respectively.  The $I$-band profile is plotted to
40 kpc while $B$ and $V$ are plotted to 35 kpc.

Lower panel:
The axisymmetrized $B-V$ and $B-I$ color profiles of NGC 4123, in 
magnitudes arcsec$^{-2}$.  The profiles have not been adjusted for 
internal reddening.
\label{fig-sbprofile}
\endfig

The luminosity profile has a strong peak at the center, a shoulder in the 
bar region, and an approximately exponential decrease outside the bar, 
though the slope changes somewhat.  The radius at which the profile breaks 
to exponential, $\sim 50\arcsec$, or 5.5 kpc, is the same as the bar 
length.  This type of surface brightness profile, with flat and 
exponential regions, was designated type II by Freeman (1970), and is 
fairly common in both axisymmetric and barred galaxies; see for example 
the sample of Palunas \& Williams (2000).  


\begintabtwo[ht]
\begin{center}
\begin{tabular}{ccccccccc}
\tableline\tableline 

Filter & Night sky &  2$\sigma$/pixel  & \multicolumn{2}{c}{exp disk}  & 
      \multicolumn{3}{c}{magnitude within}\\
   &  brightness & $\mu_{lim}$ &  $\mu_0$    & scalelength &   $R_{25}$ & $2R_{25}$ & $3R_{25}$ \\
 &  mag/$\sq\arcsec$ & mag/$\sq\arcsec$ & mag/$\sq\arcsec$ & kpc & 
   \multicolumn{3}{c}{mag} \\
\tableline
B & 22.35 & 25.11 & 21.38 &  3.46  & 12.29 & 12.09 & 12.06 \\
V & 21.52 & 24.64 & 20.65 &  3.25  & 11.65 & 11.46 & 11.45 \\
I & 19.43 & 23.38 & 19.59 &  3.20  & 10.55 & 10.36 & 10.33 \\
\tableline
\end{tabular}
\end{center}
\caption{NGC 4123 -- LCO Photometric Data}
\label{table-photpars}
\tablecomments{
The central surface brightness $\mu_0$ in column 4 has been 
corrected for inclination, but not for internal extinction.
}
\endtabtwo

Table \ref{table-photpars} summarizes overall photometric parameters
of NGC 4123 from the LCO observations.  The nominal 
optical radius of the galaxy ($R_{25}$ in $B$) is 11.1 kpc; $R_{23.5}$ 
in $I$ is 11.3 kpc.  The exponential disk scale lengths given in Table 
\ref{table-photpars} were determined from least-squares fits between 6 and 
10 kpc, where the data are well described by an exponential.  The inward 
extrapolated central surface brightness 
in $I$ of 19.59 mag (19.44 mag when corrected for extinction) makes this 
galaxy relatively high surface brightness, but not extremely so (de Jong 
1996a; Palunas 1996).  The total $I$ magnitude within $3R_{25}$ is 10.33; 
the absolute magnitude, corrected for internal extinction, is 
$M_I = -21.57$ and luminosity $L_I = 1.72 \times 10^{10}~\lsun$.
The absolute $B$ magnitude, uncorrected for extinction, is
$M_B = -19.7$, a luminosity of $0.7L_*$.

\subsection{Outer disk profiles}

We have been able to trace the surface brightness profiles to quite large 
radii, beyond $2 R_{25}$, even to $3 R_{25}$ for $I$ (9 magnitudes below 
the sky).  At very large radii the profiles become uncertain due to a 
combination of photon noise and
errors in the sky level.  Beyond 20 kpc, the color profiles are probably 
not reliable -- in particular, the reddening trend seen in $B-I$ but not 
in $B-V$ could be due to a systematic error in the $I$ sky level.  Some 
small-scale features in the profiles are certainly real, such as the red 
bump in $B-I$ at 12--14 kpc and the bump in all three filters at 16 kpc, 
which both correspond to faint spiral arms.  There is also a spiral
arm at $\sim 24$ kpc.

The surface brightness profiles show no sign of reaching a plateau before 
they disappear into the noise, at $R \sim 30$~kpc for $B$ and $V$, and 35 
kpc for $I$, so the large extent of the galaxy is not due to 
undersubtraction of the sky.  
We constructed an extended PSF from observations of bright stars
to measure the contribution of scattered light,
which is negligible.  Scattered light outside a sharply truncated disk
would produce a surface brightness profile with an observed drop of 
4 magnitudes at the disk edge, which is clearly ruled out.

In all three filters,
the galaxy is easily visible to $R \sim 25$~kpc when displayed at
high contrast, and to greater extent when the images are smoothed
with a median filter.  Spiral arms are traceable to $R \sim 25$~kpc;
the arms are strongest in the $B$ filter, while in the $I$ filter 
the extended galaxy light is more smoothly distributed than in
$B$ and $V$.

There is no evidence in the data that the 
disk is truncated at large radius,
in sharp contrast with results that suggest disk galaxies are 
generally truncated at 3 to 5 scale lengths (\eg\ van der Kruit \& Searle 
1981a,b, 1982; van der Kruit 1989;  Barteldrees \& Dettmar 1994; Morrison, 
Boroson \& Harding 1994).  The SB profiles of NGC 4123 are traceable twice 
as far, to 10 scale lengths in the $I$ filter.  At least 20\% of the galaxy 
luminosity is outside $R_{25}$. 
One other galaxy, NGC 5383, definitely
does not have a truncated disk (Barton \& Thompson 1997).

The origin of this extended stellar disk is uncertain.
It has been suggested that truncated disks occur because 
a stability condition cuts off star formation
below a critical gas surface density (Kennicutt 1989;
Sellwood \& Balbus 1999), and it is possible that stars from
the inner disk could be scattered to large radius by a
bar-like perturbation (Schwarz 1984).
However, NGC 4123 has faint spiral arms extending
to $\sim 25$~kpc, with small
blue knots that are likely sites of massive star formation.
These star-forming regions are similar to
the \hii\ regions in faint spiral arms
that Ferguson \etal\ (1998) detected to $2R_{25}$ and beyond,
in deep \ha\ imaging of three late-type spirals.
Thus at least some of the extended disk of NGC 4123
formed at large radius, although massive star formation 
in the outer disk appears to be confined to 
spiral arms, as discussed by Ferguson \etal\ (1998).
The blue knots have luminosities from $M_B = -12.5$ to at least
2 magnitudes fainter, and mean colors $B-V \sim 0.1$, $V-I \sim 0.0$.
The very blue colors indicate that the knots
are younger than $10^7$ yr (Bruzual \& Charlot 1993).  
The largest knot has $L_B = 10^7$ \lsun,
suggesting $M \sim 2 \times 10^5$ \msun.  The star formation
rate implied outside $R_{25}$ is consistent with the range found
by Ferguson \etal\ (1998).

\section{Fabry-Perot Imaging Spectroscopy}
\label{sec-fpobs}

\subsection{Observations}

We observed NGC 4123 in \ha\ emission with the Rutgers Imaging Fabry-Perot
Interferometer to obtain a 2--D velocity field with high spatial resolution.
The observations were made at the Cassegrain focus of the CTIO Blanco 4-meter
telescope, on the night of 1994 April 3.  
We used a Tektronix 1K x 1K CCD detector, with 0.36\arcsec\ pixels and a 
field of view of 160\arcsec.  
The Fabry-Perot etalon has spectral resolution of 2.5 \AA\ at \ha,
yielding a velocity resolution of FWHM $\simeq 150$~\kms.  
The instrumental spectral profile is very well approximated by a Voigt 
function, with Gaussian $\sigma_G = 26$~\kms and Lorentzian 
$\sigma_L = 33$~\kms.
The free spectral range of this etalon is 85 \AA\ at this wavelength, 
and a blocking filter of 75 \AA\ FWHM ensured that only one order was 
transmitted.

The Fabry-Perot instrument provides a normal spatial image of the sky,
but with a very narrow, tunable, bandpass.
The image is not monochromatic, but varies quadratically with radius
from the optical axis of the instrument near the center of the image,
with a 6 \AA\ gradient.
There are two temporal variations in the instrument: the wavelength 
zeropoint of the etalon varies due to temporal drift of the 
control electronics, and the position of the optical axis changes due 
to spectrograph flexure.
In practice, these complications cause little difficulty;
we take hourly calibration exposures while observing to monitor the 
drifts, and the data analysis procedures account for them and for the
radial wavelength gradient.

The observations of NGC 4123 are 18 10-minute exposures spanning
a range of 20 \AA\ in steps of approximately 1.2 \AA\ (54 \kms).
The images were processed using IRAF, and each flatfielded with a 
dome flat taken at the appropriate wavelength.
The images were then cleaned of cosmic ray events, shifted into
spatial register, convolved to a common PSF of 1.5\arcsec\ FWHM,
and sky subtracted to construct a data cube.
At each pixel in the image we smoothed the data cube spatially with a small 
Gaussian kernel of 0.6\arcsec\ FWHM, and then fitted
a Voigt function along the spectral axis.
The fits yielded maps of the velocity, line strength,
continuum level, intrinsic line width, and estimates of the uncertainties
in these quantities.
Points with intensities below a relatively small threshold or with velocity
uncertainties greater than 12 \kms\ are deleted from the maps.

The velocity map of NGC 4123 produced from the Fabry-Perot observations is 
presented in Figure \ref{fig-fpvelmap}.  The orientation is the same as 
that of Figure \ref{fig-iimage}, although the scale is different.  The 
colormap indicates heliocentric velocity; the approaching side of the 
galaxy is blue, and the receding side is red.  The diagonal tick marks
indicate the minor axis of the disk.  Assuming that the spiral 
arms are trailing, the northeast side of the galaxy is the side
closer to us.  The 
160\arcsec\ field of view does not extend much outside the bar region.  
The velocity map from 21-cm observations, presented in section 
\ref{sec-vlaobs}, extends to much greater galactic radius.

\subsection{The streaming motions in the bar}

\beginfigtwo[t]
\begin{center}
\includegraphics[width=5.75truein]{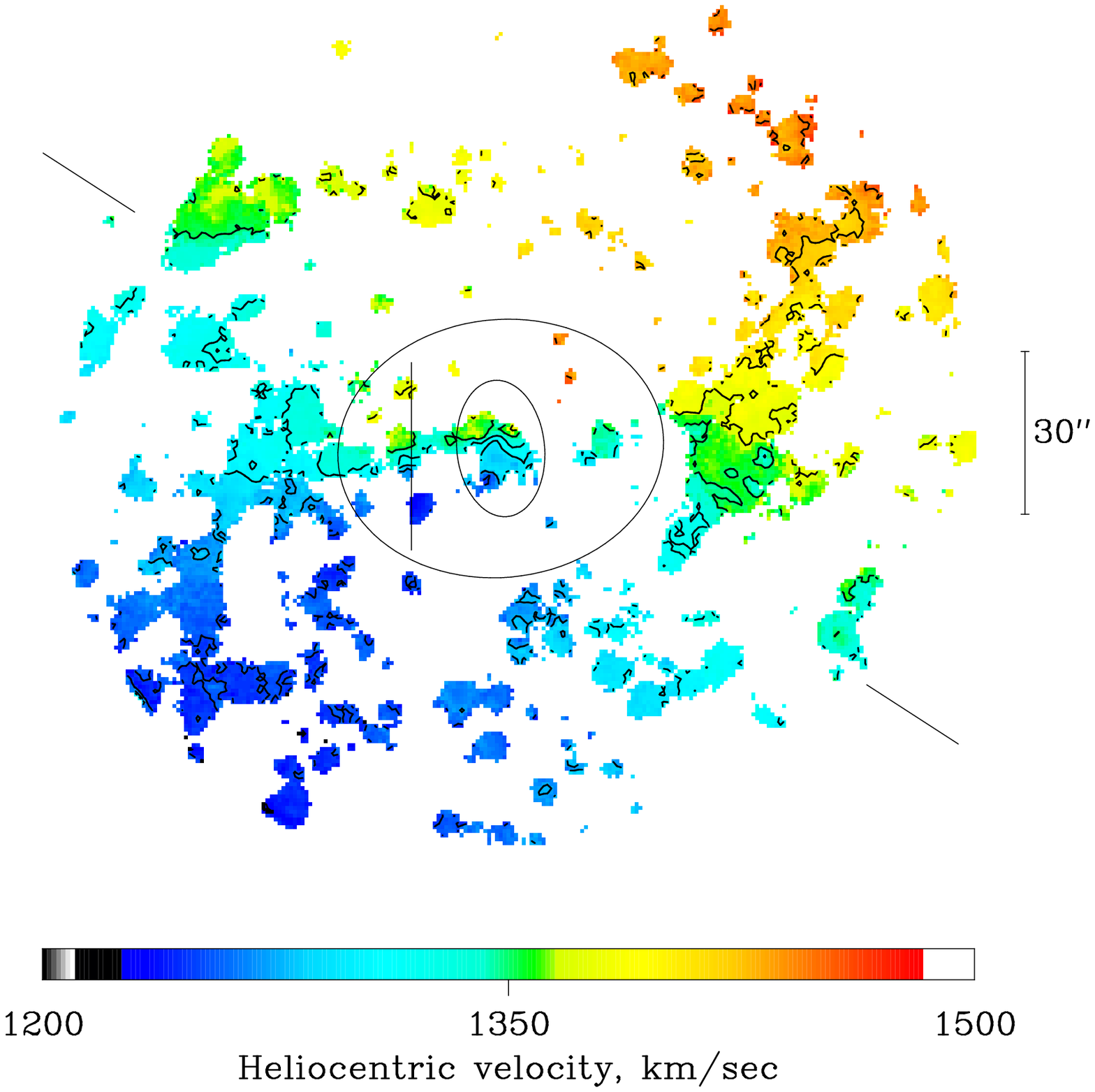}
\end{center}

\caption{The radial velocity field of the bar region of NGC 4123}
\captsize
Velocity map of the inner regions of NGC 4123, from Fabry-Perot 
observations of the \ha\ emission line.  North is up and west is to the 
right.  The area shown is 170\arcsec\ diameter,
smaller than that of Figure \ref{fig-iimage}.  
Velocity contours are plotted over the colormap at
intervals of 25 \kms.  The vertical line to 
the east of the center indicates the cut through the data plotted in 
Figure~\ref{fig-slitdat} and in Paper 
II; the data between the two ellipses are used in the goodness-of-fit 
tests in Paper II.  The diagonal lines to NE and SW indicate the 
minor axis of the disk.

\label{fig-fpvelmap}
\endfigtwo

The non-axisymmetric streaming motions associated with the bar can
be seen 
in Figure \ref{fig-fpvelmap}.  In an axisymmetric galaxy, the contour of 
the systemic velocity would run straight down the minor axis, dividing the 
map into approaching and receding sides.  In NGC 4123, the gas 
streamlines elongated
along the bar cause the isovelocity contours to twist and
run more closely along the bar, producing the S-shaped isovels
characteristic of barred galaxies (Bosma 1981).
To the east and
northeast of the galaxy center, there
are receding velocities (yellow in the colormap of Figure \ref{fig-fpvelmap})
in the bar region that are on the approaching side of the minor axis, 
and vice versa to the west.  

Furthermore, the isovelocity contours are not straight along the bar, but 
are offset from the bar major axis.  The isovels twist
as they pass through the center of the bar.  For 
example, the contour of systemic velocity, blue-green in the color map of 
Figure \ref{fig-fpvelmap}, is offset toward north on the east side of the 
bar (left of center in Figure \ref{fig-fpvelmap}), makes a jog as it 
passes through the center, and is offset toward the south on the west side 
of the bar (right of center in Figure \ref{fig-iimage}).  The regions of 
strong \ha\ emission in the bar are in similarly offset straight lines.  
There is also a sharp jump in velocity across the bar, i.e.\ parallel to 
the minor axis of the bar.  These features suggest the presence of shocks 
in the gas flow pattern, at the location of the dust lanes in the bar.

Dust lanes are presumably caused by a build-up of gas at shocks in 
the gas flow, as first suggested by Prendergast (1966)
and by many subsequent authors (\eg\ van Albada \& 
Sanders 1982; Prendergast 1983; Athanassoula 1992b; but 
cf. Beck \etal\ 1999). 
These shocks are thought to be caused by the 
non-circular gas motions along the bar:
gas falling down along the bar 
potential reaches high velocities, and as the gas climbs
away from the center, up the potential, it decelerates.
Eventually a pile-up occurs and the shock is formed.
Velocity jumps associated with bar dust lanes have been observed, in \eg\ 
NGC 5383 (Duval \& Athanassoula 1983), NGC 6221 (Pence 
\& Blackman 1984a), NGC 1365 (Lindblad \etal\ 1996a),
and NGC 1530 (Regan, Vogel \& Teuben 1997).  

The regions of strong \ha\ emission aligned with the dust lanes in
NGC 4123 suggest that compression 
of the gas by the shocks triggers star formation.  The offset of the dust 
lanes and the associated jog of the velocity contours as they pass through 
the galaxy center are evidence for an inner Lindblad resonance (ILR) at 
the bar center (Athanassoula 1992a,b).  An ILR can be inferred when some 
stellar orbits near the center are elongated perpendicular to the bar, 
rather than along it; the gas streamlines twist to follow these stellar 
orbits (\eg\ Figure 2 of Athanassoula 1992b).  This twist creates the jog 
in the velocity contours and the offset in the shocks along the bar.

The sharp jump in velocity across the bar of NGC 4123, plotted in the upper
panel of Figure~\ref{fig-slitdat}, is strong kinematic evidence 
that there are indeed shocks in the bar.  The radial velocity in the \hii\ 
regions about 20\arcsec\ east of the galaxy center changes by 80 \kms\ in 
just 4\arcsec.  
The jump is 115 \kms\ when corrected for the galaxy inclination
-- and this is just in the line-of-sight 
component of the velocity; the absolute velocity jump must be larger. 
There are also large velocity gradients in the galactic nucleus,
inside the ILR, typically 
40--50 \kms\ in observed velocity (55--70 \kms\ after inclination 
correction) across just 2\arcsec.  This angular size is barely resolved by 
these observations and the actual gradient is likely to be even steeper.
The velocity gradient shown in Figure~\ref{fig-slitdat}
is at the same location as the dust lane,
confirming that the shock and dust lane are associated.  

Because the shocks are closely linked to the non-circular gas motions,
their properties are dependent on the mass of the bar, which controls
the ellipticity of the potential.  For this reason, the strength
and location of the shocks will prove crucial for measuring the 
\ml\ and pattern speed, discussed in Paper II.

There is \ha\ emission at distances from the galaxy's center of less
than the bar radius, but not within the bar itself -- for example, the
small \hii\ regions some 20\arcsec\ northwest of the galaxy center,
and the large \hii\ region 20\arcsec\ southeast of the center.  These
regions have large velocities relative to systemic, shown in
Figure~\ref{fig-slitdat}, and provide useful data points for
constraining the models of Paper II.

\beginfig[t]
\begin{center}
\includegraphics[width=3.5truein]{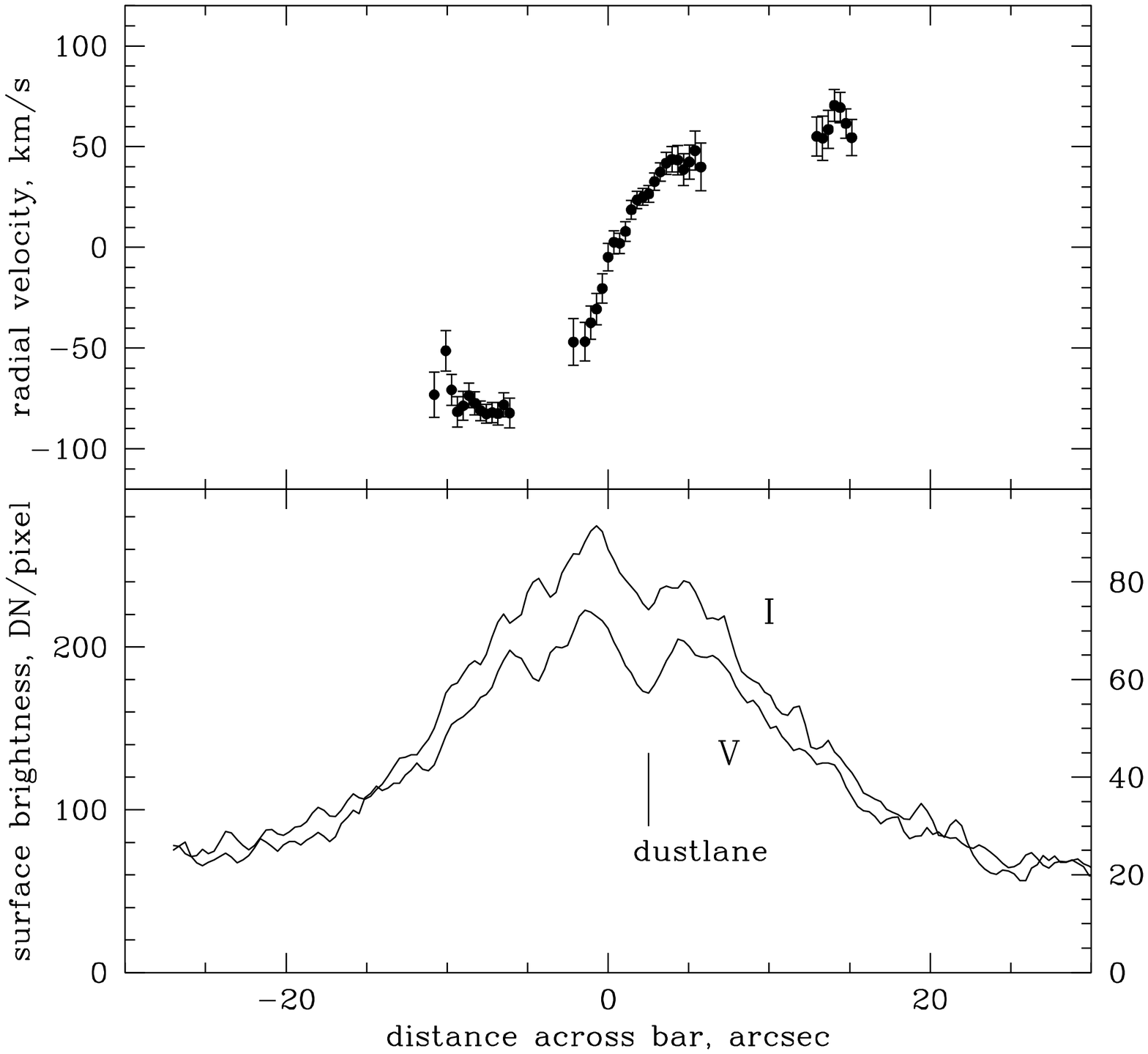}
\end{center}

\caption{Radial velocity and surface brightness profiles crossing the bar}
\captsize
The top panel shows line-of-sight velocity for the north-south
line across the bar indicated in Figure~\ref{fig-fpvelmap}.  
South is at negative $x$; the direction
of rotation of the galaxy is from left to right.
The bottom panel shows $V$ and $I$ surface brightness profiles
along the same line. The surface brightness
profiles have been sky-subtracted and smoothed with a 3 pixel
(1.1\arcsec) boxcar filter.
The units are linear
(counts/pixel) and the $V$ profile has been scaled up by a factor of 3.
The position of the prominent dustlane is indicated.
\label{fig-slitdat}
\endfig

Outside the bar radius, the non-axisymmetric motions fall off rather 
quickly and the contour of systemic velocity becomes more closely aligned 
with the minor axis.  There are spiral arms with many \hii\ regions 
visible to the northwest and southeast, but the velocity perturbations 
associated with them are much smaller than those in the bar.  The limited 
coverage and field of view of the \ha\ velocity map makes the global 
axisymmetric velocity field more difficult to pick out; the 21 cm \hi\ 
velocity map, presented below, shows it well.

\section{Radio interferometry}
\label{sec-vlaobs}

We have used the VLA to image the neutral hydrogen in NGC 4123.
The \hi\ data are complementary to the \ha\ data.  The higher angular
resolution of the Fabry-Perot data can trace the kinematics
within the bar, while the VLA data can image diffuse \hi\
and trace the rotation curve far beyond the nominal optical radius $R_{25}$.

\subsection{Observations}

We observed NGC 4123 in the \hi\ 21 cm line
with the 3km (C) and 1km (D) array of 
the Very Large Array (VLA) of the National 
Radio Astronomy Observatory\footnote{
The NRAO is operated by Associated Universities, Inc., under a
cooperative agreement with the National Science Foundation.},
for 8 and 2 hours in December 1994 and April 1995 respectively.
This combination gives an angular resolution of $\sim 15$\arcsec,
and provided 35 independent beams along the galaxy major axis,
with sensitivity to smooth structures up to 12\arcmin.
We used 63 channels with a channel width of 10.5 \kms.
Standard AIPS calibration procedures were used and the UV data
of the two arrays combined.  We subtracted the 
continuum in the $u-v$ domain using UVLIN, using 19
line-free channels.  

We imaged the data with the AIPS
task IMAGR using robust weighting (Briggs 1995),
yielding a sensitivity almost equal to natural weighting and 
a resolution almost equal to uniform weighting.
An elliptical Gaussian fit to the beam 
yielded a beam shape of 19.1\arcsec\ x 16.1\arcsec\ FWHM with a
position angle of 56.5\deg.
The rms noise in line-free channels after
continuum subtraction is 0.38 mJy/beam.
We cleaned the images using the CLEAN algorithm
to a flux limit of 0.19 mJy/beam, and made images of total \hi\
intensity, the intensity weighted velocity field, and 
the velocity dispersion with the AIPS task MOMNT.  
The total 21 cm flux observed is 63.5 Jy~\kms,
which recovers all the single-dish flux as tabulated
in Huchtmeier \& Richter (1989).

\beginfigtwo[t]

\begin{center}
\includegraphics[angle=-90,width=6.5truein]{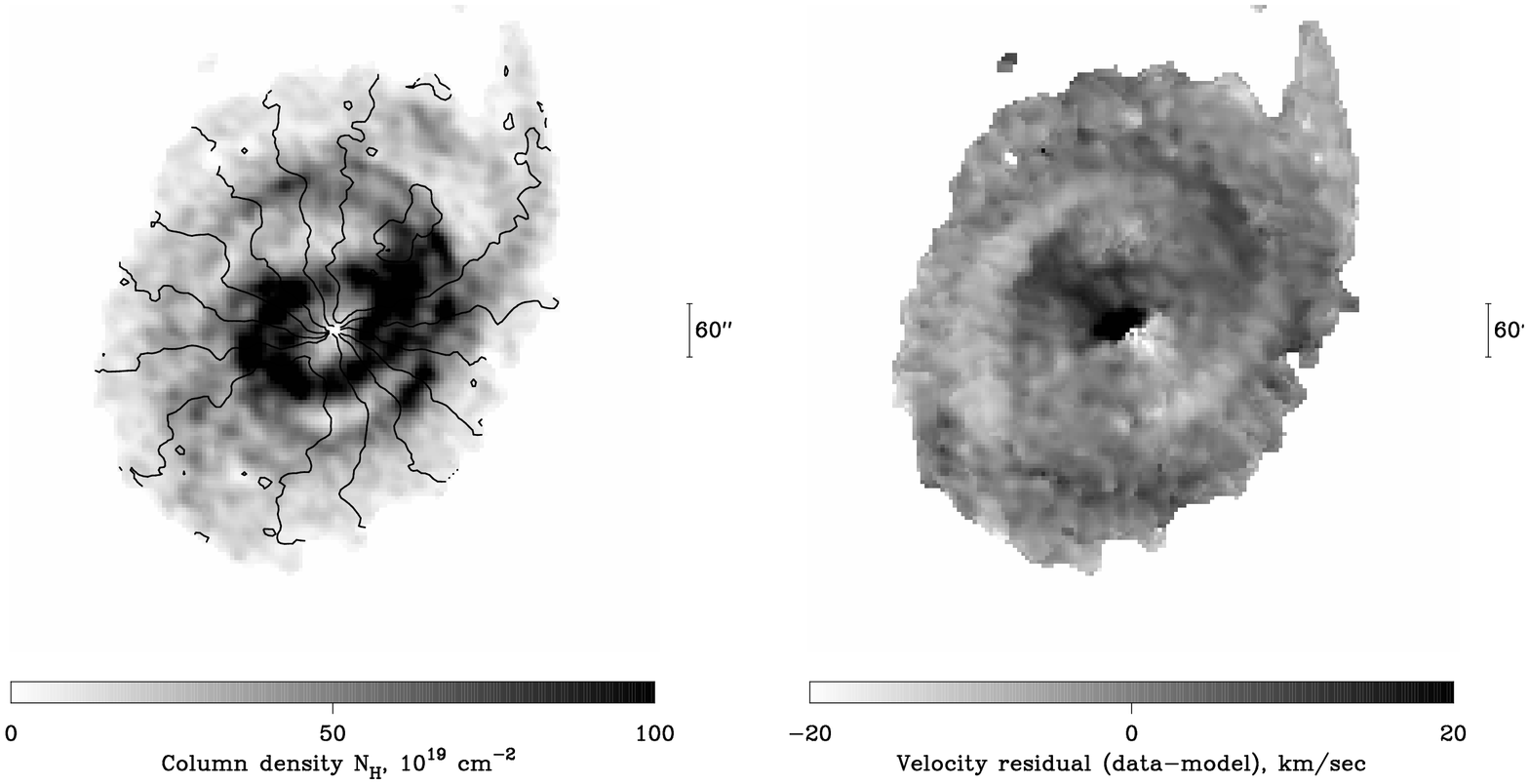}
\end{center}

\caption{21 cm emission, velocity field, and residual velocity}
\captsize
Left panel:
The 21 cm surface brightness is plotted 
as grayscale, with velocity contours overlaid.  North is up
and west is to the right.  The surface brightness
grayscale runs from $N_H = 0$ to $10^{21}$ \percmsq.  
The contours are of heliocentric velocity from 1225 to 1425 \kms\
at intervals of 25 \kms, with receding side in the northwest.\\
Right panel:
The residuals from our rotation curve extraction are
plotted (data -- model), on a grayscale from
--20 to 20 \kms.  Non-circular motions are
seen in the bar and in the outer spiral pattern.

\label{fig-vlamap}
\endfigtwo


\subsection{Discussion}

The VLA 21 cm intensity and velocity field shown in the left panel of 
Figure \ref{fig-vlamap} reveals that NGC 4123 has a large \hi\ disk 
of 11\arcmin\ x 7.5\arcmin, and a regular velocity field.
The highest \hi\ column densities occur in a ring where the surface 
brightness is mostly above $10^{21}$ \percmsq.
The ring's radius is some 
70\arcsec, or 7.6 kpc, placing it somewhat outside the bar.  
A loose spiral pattern is also visible outside the ring.  
Inside the ring, 
the surface brightness decreases toward the center, from 
$5 \times 10^{20}$ \percmsq\
to a low of $4 \times 10^{19}$ \percmsq\
at the center.  The hole in the center of the 
gas distribution is frequently seen in barred galaxies; \eg\ NGC 1300 
(England 1989) and NGC 1398 (Moore \& Gottesman 1995).  It is presumably 
caused by the bar, which sweeps gas inside its corotation radius from the 
bar region towards the galaxy center, as seen in fluid-dynamical 
simulations (\eg\ Athanassoula 1992b; Piner \etal\ 1995; Weiner \& 
Sellwood 1999; Paper II).

The overall velocity field in Figure \ref{fig-vlamap} is quite regular.  
The effect of the bar is visible only inside a radius of $\sim 30$\arcsec; 
the contours close to the systemic velocity in this region are twisted to 
run more east-west, along the bar, rather than along the minor axis of the 
galaxy.

In the outer parts of the \hi\ disk, velocity perturbations due to the 
spiral arms are visible as kinks in the velocity contours.  These are 
particularly noticeable in the velocity contours about 160\arcsec\ 
northeast of the galaxy center.  The contours bow to the northwest as 
they cross the spiral arm.  The effect is of order 10--20 \kms\ in 
line-of-sight velocity.  Similar spiral-arm-related kinks of lesser 
magnitude are visible in other regions of the velocity field.

\beginfigtwo[t]
\begin{center}
\includegraphics[width=5.5truein,angle=-90]{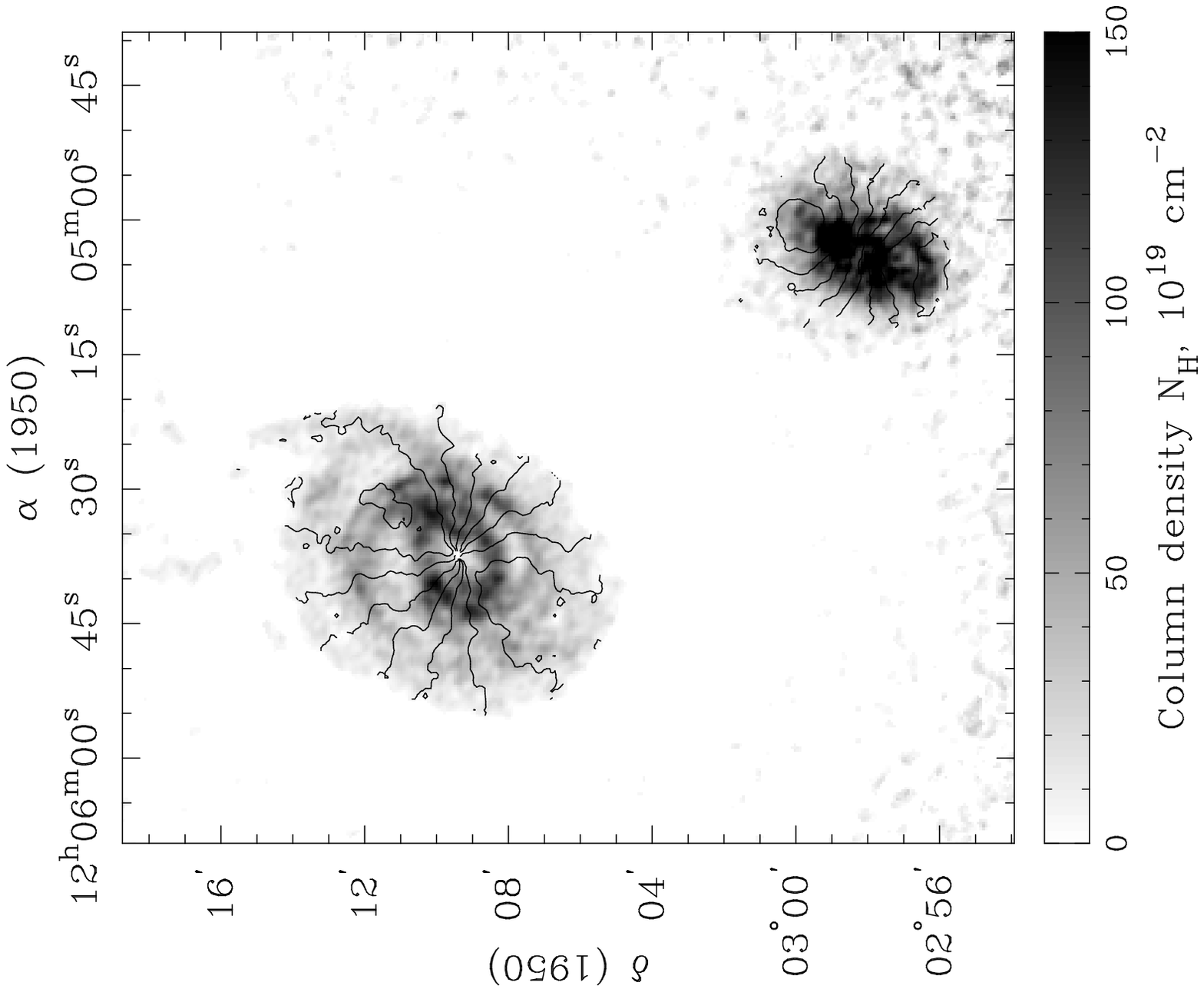}
\end{center}

\caption{VLA map of NGC 4123 and NGC 4116}
\captsize
The 21 cm surface brightness of NGC 4123 and NGC 4116 is plotted 
as grayscale, with velocity contours overlaid.  North is up
and west is to the right.  The noise increases in the south and west
as these regions are at the edge of the VLA primary beam.
The surface brightness
grayscale runs from $N_H = 0$ to $1.5 \times 10^{21}$ \percmsq.  
The contours are of heliocentric velocity from 1225 to 1425 \kms\
at intervals of 25 \kms, and each galaxy has
the receding side in the northwest. 
\label{fig-twogals}
\endfigtwo

The only other unusual features in the \hi\ distribution are the 
low-column-density arm or extension on the northwest side of the disk, and 
the even lower column-density fragments of \hi\ emission at the north of 
the disk.  The shape of these features suggests that they could be due to 
some kind of tidal interaction, although because they are close to the 
detection limit, the ``fragments'' may not be as isolated as they appear.  
In any case, the kinematics of the arm and fragments are completely 
consistent with unperturbed circular rotation.  There is no obvious 
companion very close to the arm, either in the Palomar Sky Survey, the 
Digitized Sky Survey, our $BVI$ imaging, or the 21 cm observations.  As 
mentioned in the introduction, the galaxy NGC 4116 is just 14\arcmin\ to 
the southwest of NGC 4123 and is also visible in our VLA observations.  We 
included it in the imaging and CLEANing process and mapped its \hi\ 
emission and velocity.  Figure~\ref{fig-twogals} shows the total
\hi\ image and velocity field of
both galaxies, corrected for the primary beam response.
NGC 4116 has an \hi\ disk just over half the angular size of the 
\hi\ disk of NGC 
4123, and kinematics consistent with circular rotation and a 
slowly rising rotation curve.

Tides from NGC 4116 could be responsible for the arm of NGC 4123, but if 
this is the explanation, it is curious that the counter-arm pointing away 
from the companion is the more pronounced; furthermore, the larger NGC 
4123 does not appear to have distorted the smaller galaxy.  
The kinematics of the arm can be modeled 
in circular rotation with the rest of the \hi\ disk of NGC 4123, as 
described below.  Moreover, there is no sign that NGC 4116 has any 
significant influence on the internal dynamics of NGC 4123.

\subsection{The rotation curve of NGC 4123}

We derived a rotation curve from the \hi\ data for NGC 4123 using a 
program written by Palunas (1996) which models the galaxy as a 
circular, differentially rotating flat disk, and assumes no {\it a priori} 
form for the shape of the rotation curve.  The program fits
the velocity data in a series of elliptical annuli by adjusting several
parameters: central position $x_0,y_0$, position angle $\phi$, 
inclination $i$,
and rotation speed $V_c$ within each annulus.  Data near the minor axis
are excluded since they have almost no signal due to projection.
In practice the kinematic center is not well constrained 
(see also Begeman 1987, Palunas 1996) and 
we fixed the central position to be the photometric center of
the galaxy.

Rotation curve fits to two-dimensional velocity fields 
have degeneracies between the parameters, and are influenced by
non-circular motions in the disk and warps.  We examined
the variation of $\phi$, $i$, and $V_c$ with radius.  Inside
the bar, the parameters are ``twisted'' by the non-circular
motions in the bar.  Outside the bar radius, from $\sim 130\arcsec$
to 250\arcsec, $\phi$, $i$, and $V_c$ change only slowly and
we adopted median values of $\phi$ and $i$ from this region,
yielding  $\phi = -33 \pm 2\deg$ and $i=45 \pm 4\deg$.  
$V_c(R)$ and $i(R)$ are covariant and show artificial excursions 
at certain radii, due to spiral arms crossing the
annuli (see Palunas 1996).  However, by averaging over a radial range 
we obtain a reliable estimate of the parameters $\phi$ and $i$.

We then extract the rotation curve $V_c(R)$ in annuli
using the adopted values of $x_0$, $y_0$, $\phi$, and $i$.
We chose to use global values of $\phi$ and $i$ rather 
than doing a tilted-ring extraction, since tilted-ring fits
do not work well on galaxies of $i \sim 45\deg$ (\eg\ Begeman 1987)
and are also thrown off by spiral arms.  
The rotation curve extraction was spatially oversampled, in part
to avoid further smoothing caused by binning into annuli.
The residual map in the right panel of Figure \ref{fig-vlamap}
shows that the model produces a good representation
of the data, justifying the use of global $\phi$ and $i$.
Outside the bar region, residuals are generally below 10 \kms.

\beginfigtwo[ht]
\begin{center}
\includegraphics[width=5.25truein]{fig8.ps}
\end{center}

\caption{Mass models for the rotation curve of NGC 4123}
\captsize
The contributions of the stellar disk, gas disk, and best-fit halo to the 
rotation curve of NGC 4123 are shown for models with low and high mass 
disks, and three different dark halo types. \\
Points and errorbars -- observed rotation curve. \\
Uppermost solid line -- best-fit total rotation curve. \\
Dotted line -- contribution of stars+gas to the rotation curve. \\
Lower solid line -- contribution of the stellar disk. \\
Dashed line -- contribution of the best-fit halo. \\
Dot-dashed line -- contribution of the gas disk. \\
The left column has disk $\ml=1.0$, and the right column
has disk $\ml=2.25$.  The top row uses isothermal halos,
the middle row uses NFW halos, and the bottom row uses 
Moore-type halos.
\label{fig-massmodels}
\endfigtwo

The rotation curve computed from the 
\hi\ data is shown as the points and errorbars in each panel of Figure 
\ref{fig-massmodels}.  The approaching and receding sides of the galaxy 
are plotted separately.  The rotation curves 
for the two sides differ by 10--20~\kms\ at large 
radii; the difference is
caused by spiral non-circular streaming motions.  The right
panel of Figure \ref{fig-vlamap} shows the residual velocity field 
after subtraction of the axisymmetric rotation curve 
(using an average of the two sides; the residual map is similar
if the approaching and receding sides are subtracted separately).
There are strong antisymmetric residuals in the bar region, and a 
spiral-shaped residual from the outer spiral pattern.  

Outside the bar radius of 5.6 kpc, it is reasonable to derive a rotation 
curve assuming that the gas is on circular orbits.  Inside that radius, 
the derived rotation curve is affected by noncircular motions and the 
error bars increase because the data are not fully consistent with 
circular orbits and because the \hi\ disk has a central hole.  

The maximum of the rotation curve is 145 \kms, while the 
flat part of the rotation curve is at $V_c = 130~\kms$.
Comparing the rotation 
width and total $I$ magnitude to the Tully-Fisher relation of Giovanelli 
\etal\ (1997) suggests that the galaxy is overluminous by 0.2 magnitudes, 
which is consistent with the scatter in that TF relation.  The difference 
could also be caused by a 10\% overestimation of the distance to NGC 
4123, which would require a 10\% increase to the \ml\ values we discuss 
below.  

\subsection{The spiral pattern speed}

The spiral velocity residual is a one-armed
pattern that changes sign several times along the arm.  An $m=1$
velocity residual is expected inside corotation of an $m=2$ spiral
pattern (Canzian 1993), but an $m=3$ pattern is expected outside
corotation, which suggests that nearly all of the \hi\ disk of NGC 4123 
is inside corotation of the spiral.
However, the $m=1$ spiral residual is not exactly as in Canzian's
model, where the sign of the residual does not change along the arm.
Alternatively, if the residual is interpreted as a streaming motion,
it is consistent with inward streaming along the arm until the
the last sign change, at $200\arcsec$ along the major axis
southeast of the galaxy center.  Past this point the positive 
residual on the approaching, far side of the galaxy indicates 
outward streaming.  The change from inward to outward streaming 
occurs at the corotation radius, which is then at 21 kpc.  Again
most of the \hi\ disk is inside corotation.

In Paper II, we show that the corotation radius of the bar is 
well within the \hi\ disk, at 5 to 9 kpc.  Thus the bar and spiral
pattern have different corotation radii and different pattern
speeds, even though the spiral arms appear to emerge from the 
ends of the bar.  These dynamically determined pattern speeds 
confirm the suggestion of Sellwood \& Sparke (1988) that bar
and spiral pattern speeds in the same galaxy can be significantly
different.

\section{Mass models}
\label{sec-obsmassmodels}

\subsection{Fitting the rotation curve}

We fit axisymmetric mass models made up of the stellar disk, gas disk, and 
a dark halo to the rotation curve.  Because they are axisymmetric, these 
models neglect the bar distortions both in the light and in the velocity 
field, which are strong only inside the bar radius of 5.6 kpc.  
Our goal is to determine the halo 
required to fit the outer rotation curve for a range of values for the 
disk mass-to-light ratio.  We also want to determine the maximum disk 
mass-to-light ratio, 
the highest disk \ml\ consistent with the rotation 
curve data and a non-hollow dark halo.

We calculate the rotation curve of the axisymmetrized stellar disk from 
the $I$-band image, assuming a uniform \ml.  Our procedure is described in 
more detail in Paper II.  Briefly, we remove foreground stars and rectify 
the image to face-on, using the position angle and inclination obtained 
from the \hi\ observations.  We then average the image in circular annuli 
about the center to produce an axisymmetrized surface brightness 
distribution.  We assume that the stellar disk has a vertical distribution 
of the common form $\rho(z) \propto {\rm sech}^2 (z/z_0)$ and a scale 
height of $z_0 = 400$~pc, similar to that of the disk of the Milky Way 
(Mihalas \& Binney 1981).  Following Quillen \etal\ (1994), we calculate 
the gravitational accelerations as a function of position, summing the 
contributions of each pixel efficiently using FFT techniques (Hockney 
1965).  The acceleration as a function of radius determines the stellar
contribution $V_{disk}(R)$ to the rotation curve $V_c(R)$.

The rotation curve due to the stellar disk for an assumed \ml\
is plotted as the lower solid 
line in the panels of Figure \ref{fig-massmodels}.  It is plotted out to 
the end of the optical image only, though its contribution is accounted 
for outside this radius.  Its slightly unusual shape, rising quickly, 
leveling off, then rising again, is due to the shoulder in the 
axisymmetrized surface brightness profile (Figure \ref{fig-sbprofile}) in 
the bar region.

We calculate the rotation curve due to the gas disk in a similar fashion.  
We assume that the \hi\ disk is thin and flat and rectify it to face-on, 
average the column density azimuthally, and scale the mass
up by 1.4 to correct for helium.  
The atomic gas contribution to the rotation curve, shown as 
the dot-dashed line in Figure \ref{fig-massmodels}, is everywhere a rather 
small fraction of the observed velocity, and inside 7 kpc it makes a 
very small negative contribution to ${V_c}^2$ 
(i.e. outward centripetal acceleration) due to the central hole in the 
atomic gas density.  The dotted line in Figure \ref{fig-massmodels} 
is the combined 
contribution of the stellar and gas disks -- the 
contribution of visible, non-dark matter to the rotation curve.

We do not have CO imaging data for NGC 4123, and thus cannot make
a detailed calculation of the contribution of its molecular gas
to the rotation curve.
Single-dish CO observations with the SEST and IRAM 30-m
imply a molecular gas mass of 
$4.0 \times 10^8~\msun$ within a 23\arcsec\ beam, and
$5.1 \times 10^8~\msun$ within a 45\arcsec\ beam
(Chini, Kruegel \& Steppe 1992).  The contribution to
the rotation curve from the molecular gas is then 
37~\kms\ at 1.25 kpc and 30~\kms\ at 2.5 kpc.
The molecular gas may at most fill in the 
hole in the atomic gas distribution, but its contribution 
to the overall rotation curve is negligible.

The dark halo needed to fit the observed rotation curve is specified by 
two parameters: a characteristic density and radius.  We performed
least-squares fits for these parameters over a range of disk \ml\
and three different models for the dark 
halo: a pseudo-isothermal model with a constant density core and $r^{-2}$ 
envelope:
\begin{equation}
\rho_{\rm iso}(r) = \rho_0 \frac{1}{1 + (r/r_c)^2}
\end{equation}
and two broken power-law models, the ``NFW profile'' 
and ``Moore profile'' found in 
$N$-body simulations of dark matter halo formation (Navarro, Frenk, \& 
White 1996; Syer \& White 1998; Moore \etal\ 1999).  
The NFW-type model has an inner 
$r^{-1}$ and outer $r^{-3}$ dependence, and break radius $r_s$:
\begin{equation}
\rho_{\rm NFW}(r) = \rho_s \frac{4{r_s}^3}{r(r+r_s)^2}.
\end{equation}
This pseudo-NFW halo model is 
not precisely an ``NFW model,'' because the NFW family of halos is a 
one-para\-meter family: Navarro, Frenk \& White (1996) found a relation 
between the characteristic density and radius for halos formed in any 
given cosmology, whereas we allow these parameters to vary independently 
in order to fit the rotation curve.  An NFW-type profile can 
equivalently be uniquely expressed as a function of concentration 
index $c$ and virial velocity $V_{200}$ (Navarro \etal\ 1996).
The Moore-type profile is similar to the NFW halo but has 
an inner cusp slope of $r^{-1.5}$:
\begin{equation}
\rho_{\rm Moore}(r) = 
  \rho_s \frac{2}{{r/r_s}^{1.5} \left(1 + (r/r_s)^{1.5}\right)}.
\end{equation}
Halos with this profile are found in high-resolution
$N$-body dark matter simulations (\eg\ Moore \etal\ 1999).

These expressions 
specify spherical mass distributions, although the rotation curve 
constrains only the
centripetal acceleration in the disk plane; for any degree of halo
flattening, a volume density distribution could be devised to 
yield the same rotation curve.  The functional forms used are 
analytically convenient, but are not intended to favor specific
models for halo formation; we have also not attempted to modify
the halo profiles for compression by disk collapse, a point
we discuss further in Paper II.

The fits are not very sensitive to the functional form adopted: 
all three types of halo model fit the data well for disk $\ml \leq 2.25$,
although for high-mass disks with $\ml \geq 2.0$, the cuspy
NFW and Moore halos are forced to very low scale densities and 
large scale radii in order to contribute to the rotation
curve at large radius but avoid exceeding the observations
at $R \sim 10 kpc$.

We constructed models over the range $\ml = 1.0$ to 3.0 with 
the three halo profiles.  The models and best-fit halo 
parameters are listed in Table \ref{table-halopars}.  For all
profile types, the fitted halo 
density and radius are strongly covariant in a way which keeps the halo 
contribution to the outer rotation curve roughly constant, although
the actual error ellipse on the fitted parameters is quite narrow -- 
an example is shown in Paper II.  
We use the halos fitted in these mass models in the 
fluid dynamical simulations described in Paper II.

Columns 7 and 8 of Table \ref{table-halopars}
show the results of fitting NFW-type profiles by fitting for
$c$ and $V_{200}$.  The best-fit halo profile $\rho(r)$ is the same
as that obtained by fitting for $\rho_s$ and $r_s$,
but $c$ and $V_{200}$ are much less covariant than $\rho_s$ and $r_s$
and are tightly constrained for a given \ml.
As disk \ml\ is increased, the halo contribution inside the 
optical radius decreases greatly and the radius at which the
halo rotation curve peaks moves out, requiring a lower 
concentration index $c$.  In order to maintain the necessary
halo contribution at large $R$, the virial velocity $V_{200}$
must increase as $c$ decreases; the 
virial radius $r_{200}$ increases with $V_{200}$.
For $\ml \geq 2.0$, the halo rotation curve peaks beyond
the end of the \hi\ data at 47 kpc.


\begintabtwo[ht]

\begin{center}
\begin{tabular}{crrrrrrrrrrr}
\tableline\tableline 

Disk &  \multicolumn{3}{c}{Isothermal halo} &  
             \multicolumn{3}{c}{NFW-type halo} &
             \multicolumn{2}{c}{NFW-type halo, $\rho(c,V_{200})$} & 
             \multicolumn{3}{c}{Moore-type halo} \\
$M/L_I$  &  $\rho_0$\phn  &  $r_c$\phn  &  \rchisq  &   
               $\rho_s$\phn\phn  &  $r_s$\phn  & \rchisq  &
	      $c$\phm{ 0.000} & $V_{200}$\phm{ 0.0}  &
               $\rho_s$\phn\phn  &  $r_s$\phn  & \rchisq  
              \\ \tableline

1.00 & 143.0\phn &  1.24 & 1.76  & 5.47\phn\phn & 6.82  & 0.71  & 
   $14.2\phn\phn \pm 0.09\phn$ &  $81.2 \pm 0.7$  &
   1.60\phn\phn &  12.1  &  0.76  \\
1.50 &  97.4\phn &  1.39 & 0.87  & 1.88\phn\phn & 10.5\phn & 0.72  &
   $9.2\phn\phn \pm 0.07\phn$  &  $81.0 \pm 0.8$  &
   0.522\phn    &  19.0  &  0.76  \\
1.75 &  36.1\phn &  2.24 & 0.84  & 0.92\phn\phn & 14.2\phn & 0.77  &
   $6.76\phn \pm 0.06\phn$  &  $82.6 \pm 0.9$  &
   0.237\phn    &  26.9  &  0.83  \\
2.00 &  12.7\phn &  3.76 & 0.87  & 0.39\phn\phn & 21.4\phn & 0.85  &
   $4.35\phn \pm 0.04\phn$ &  $87.5 \pm 1.0$  &
   0.0802       &  45.4  &  0.96  \\
2.25 &   4.68 &  6.33 & 0.93  & 0.13\phn\phn  & 38.2\phn & 1.02  &
   $2.09\phn \pm 0.03\phn$ &  $93.0 \pm 1.2$  &
   0.0399       &  61.0  &  1.33  \\
2.50 &   1.91 & 10.4\phn & 1.11  & 0.043\phn & 69.1\phn & 1.40  &
   $1.30\phn \pm 0.01\phn$ &  $116.3 \pm 1.7$ &
   0.0101       & 128.3  &  1.93  \\
2.75 &   0.88 & 16.9\phn & 1.63  & 0.011\phn & 185.7\phn & 2.10 &
   $0.645 \pm 0.005$ &  $142.8 \pm 2.2$ &
   0.0063       & 156.3  &  3.09  \\
3.00 &   0.46 & 27.1\phn & 2.75  & 0.0038 &381.7\phn & 3.46 &
   $0.255 \pm 0.002$ &  $151.1 \pm 2.9$ &
   0.0040       & 179.4  &  4.77  \\
\tableline
\end{tabular}
\end{center}

\caption{Mass models: best-fit halo parameters}
\label{table-halopars}
\tablecomments{
The best-fit halo parameters for disk mass-to-light ratios from 1.0 to 
3.0.  Column 1 is the disk \ml.  Columns 2, 3, and 4 are for isothermal 
halos; columns 5, 6, and 7 are for NFW-type power-law halos; and
columns 10, 11, and 12 are for ``Moore-type'' power-law halos.
The densities are in 
$10^{-3}~\mden$, and the radii are in kpc.  Columns 8 and 9
are the parameters and $1\sigma$ errors for the best fit 
NFW-type halos when the fitted parameters are the concentration 
index $c$ and virial velocity $V_{200}$.  
}
\endtabtwo

\subsection{Discussion}


We plot rotation curve decompositions for \ml\ of 1.0 and 2.25 in 
Figure \ref{fig-massmodels}.  For $\ml \leq 2.25$, all mass models fit the 
observed rotation curve well and the small differences in the values of 
reduced \chisq, given in Table \ref{table-halopars}, are not significant, 
given that the true shape of the halo radial profile is unknown and could 
be adjusted slightly to improve the fit.

Models with too high a disk \ml\ cannot fit well, however, because the 
fitted rotation curve exceeds the observed curve.  The stellar disk alone 
accounts for the entire rotation curve at 8 kpc when $\ml = 2.5$ 
and the model is an acceptable fit since it only 
marginally exceeds the data when the isothermal halo is included.  For 
higher $\ml$, the total rotation curve is 
too high, and the fit is unacceptable.  For an isothermal halo, therefore, 
the maximum disk \ml\ is about 2.5.  NFW and Moore-type halos, which are more 
centrally concentrated, contribute more to the peak of the rotation curve, 
and for them the maximum disk \ml\ is
slightly less, about 2.25.  For maximum disk models, the NFW and
Moore halos have low densities and large scale radii.

\subsection{Radial variation in \ml}

It is conventional to assume a constant \ml\ when fitting rotation
curves, but it is not guaranteed, although there is some
evidence for constant \ml\ (van der Kruit 1981).  
There is a color gradient
in NGC 4123, as discussed in Section~\ref{sec-sbprofiles}, which 
could be caused by a change in the disk stellar population.
To test the effect of a radial variation in \ml, we compared the 
stellar disk contribution to the rotation curve for a constant \ml\ model 
and for models in which the \ml\ declines outside the bar radius.
The stellar population is likely to be radially mixed inside
the bar radius, and in Paper II we will determine the stellar \ml\ inside
the bar radius, so the important question is whether the rotation
curve decomposition is strongly affected by a disk \ml\ which differs
inside and outside the bar radius.
Figure \ref{fig-mlvary} shows the disk contribution for three models:
a constant $\ml = 2.25$ model, and two models with a steep decline in \ml\
outside the bar radius.  

Even these models with a
rather exaggerated decline in \ml\ only have a small effect
on the disk contribution to the rotation curve.  Reasonable
variations in \ml\
therefore do not substantially increase the amount of dark
matter required by the rotation curve.  An increase in disk \ml\
with radius, as could happen if the outer disk is bluer because
it is old and metal-weak, would of course reduce the required 
dark matter, but again the effect will be small.

The effect of varying \ml\ is small because at 
the bar radius of 5.6 kpc, the disk contribution to the rotation curve is 
already nearing its peak.  The decline from the peak of a rotation curve 
is relatively gradual, whether or not the mass distribution is cut off.
A declining \ml\ outside 
the bar radius cannot make the rotation curve decline fast enough to 
change the inferred halo parameters significantly -- in fact, a declining 
\ml\ makes the disk part of the rotation curve peak up by a tiny
amount,  because it behaves like a disk truncation.

\beginfig[t]
\begin{center}
\includegraphics[width=3.5truein]{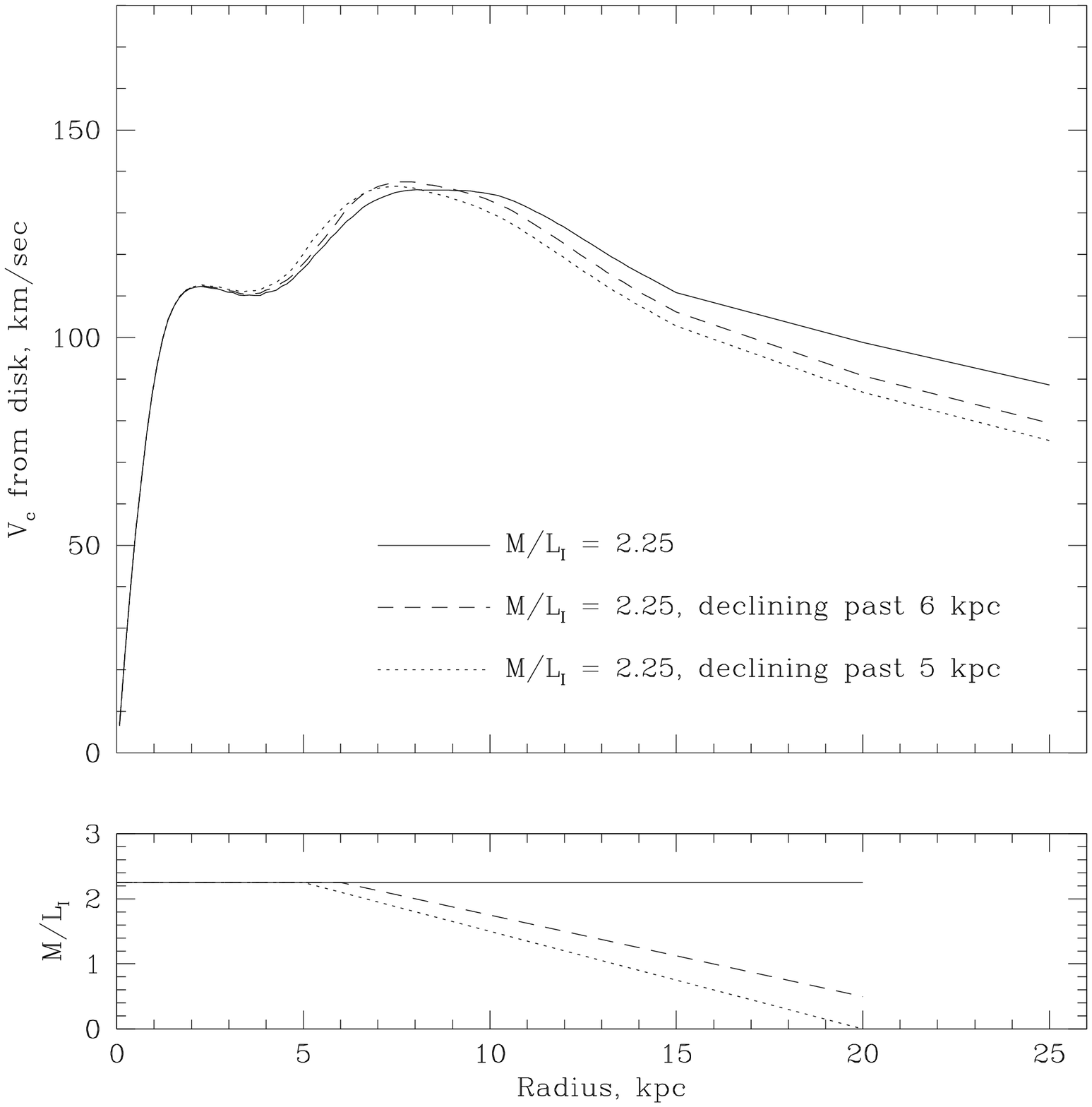}
\end{center}

\caption{Effect of varying $M/L$ on disk rotation curve}
\captsize
The disk contribution to the rotation curve is shown for
three models.  Solid line: \ml = 2.25, constant; dashed line:
\ml\ declining outside 6 kpc, to 0.5 at 20 kpc; dotted line:
\ml\ declining outside 5 kpc, to 0 at 20 kpc.
\label{fig-mlvary}
\endfig

\section{Conclusions}

NGC 4123 is a strongly barred galaxy with a rather boxy bar containing two 
components, both of which are more rectangular than ellipsoidal.  The bar 
has the straight offset dust lanes characteristic of strong bars 
(Athanassoula 1992b), although they are less obvious in the $I$ image than 
in bluer bandpasses.  The outer disk of the galaxy shows a roughly 
exponential decline of surface brightness with radius to the limit of the 
data, at $I \sim 28; B, V \sim 29$.  The disk extends to at least 10 
exponential disk scale lengths; there is no sign that it is truncated, in 
agreement with 1 mag deeper measurements of NGC 5383 (Barton \& Thompson 
1997) but in sharp contrast to the conventional result that disks are 
truncated at 3-5 scale lengths (cf. van der Kruit \& Searle 1981a,b, 1982; 
van der Kruit 1989; Barteldrees \& Dettmar 1994).  Blue knots in the
outer spiral arms indicate that stars are forming well outside
$R_{25}$.

Fabry-Perot observations of the velocity field show strong 
non-axisymmetric motions inside the bar radius, and large velocity jumps 
on the leading edges of the bar, at the location of the dust lanes.  These 
support the interpretation of the dust lanes as the location of shocks in 
the gas flow.  The velocity jumps are offset and several velocity contours 
make a sharp jog as they passes through the center of the galaxy, 
suggesting the presence of an inner Lindblad resonance.  The strength of 
the non-axisymmetric motions decreases rapidly beyond the bar radius.

21 cm observations of NGC 4123 show that it has an
extended gas disk in circular rotation outside the bar radius, with
spiral structure visible in the intensity and weakly in the
kinematics.  The kinematics indicate that corotation 
of the spiral pattern is at $\sim 4$ times
the bar radius; when combined with the result from Paper II that the
bar is fast-rotating, this implies that the bar and spiral have
very different pattern speeds.
There is no evidence for a warp.  The rotation curve
derived from these observations extends to 4 optical radii 
($4 R_{25}$) and requires the presence of a dark matter halo in addition
to the luminous disk.  The maximum height of the rotation curve is
$\simeq 145~\kms$; the height of the flat part of the rotation curve
is $130~\kms$.

Mass models that include a stellar disk and dark halo to the \hi\ 
rotation curve yield acceptable fits, in a $\chisq$ sense, for $I$-band 
disk mass-to-light ratios of 2.5 or less.  Model halos with either
isothermal or power-law density distributions fit the data well. 
The axisymmetric rotation curve does not contain enough information to 
constrain the disk-halo decomposition further.  Both heavy and light disk 
models, and both isothermal and power-law dark halos, yield acceptable 
fits to the rotation curve.

However, there is extra information in the non-circular gas streaming 
motions seen in the Fabry-Perot data.  In Paper II, we use this 
information to show that the disk of NGC 4123 is close to maximum,
with $\mli = 2.25$ the preferred value.

\acknowledgments

We thank the staff at CTIO, the VLA, and Las Campanas for their excellent 
support during these observations.  BJW thanks Karl Gebhardt and Povilas 
Palunas for Fabry-Perot reduction software and helpful comments.
We thank Jason Prochaska for a reading of the manuscript and the
anonymous referee for helpful comments.
This research was supported in part by NSF grant AST 96/17088 and NASA 
LTSA grant NAG 5-6037 to JAS, by NSF grant AST 96-19510 to TBW, 
and NSF grant AST 96-17177 to JvG.
BJW has been supported by a Barbara McClintock postdoctoral
fellowship from the Carnegie Institution of Washington.
Operation of the RFP is partially supported by CTIO.


\end{document}